\documentclass[aps,onecolumn,12pt,final, showpacs]{article}
\usepackage{amsmath,amssymb,amsfonts}
\usepackage{graphics,graphicx}
\usepackage{color}
\usepackage{verbatim}
\usepackage{latexsym}
\def \d {\mathrm{d}}
\def \exp {\mathrm{exp}}
\def \th {\textsuperscript{th}\ }	
\begin{document}

\title{Formation and Search of Large Scale Antimatter Regions}

\author{A.V. Grobov\thanks{alexey.grobov@gmail.com} \\ S.~G.~Rubin\thanks{sergeirubin@list.ru} \\National Research Nuclear University MEPhI \\ (Moscow Engineering Physics Institute)}

\date{}
\maketitle

\abstract{In the present paper we discuss a generation of large antimatter regions with sizes
exceeding the critical surviving size. In the modern epoch domains with high antimatter
density evolve to single galaxies with a peculiar content of anti-helium and anti-deuterium.}

\section{Introduction}

Astrophysical observations indicate that our local Universe is filled mostly with baryons. If the baryon charge was zero from the beginning the total baryon excess in modern epoch should be much smaller than the observed density \cite{ZeldDolg}. It means that the baryon charge has been dominating in the early Universe. The charge could be distributed uniformly throughout the Universe or some antimatter domains could exist.

Characteristic size and abundance of such antimatter islands depends on initial conditions and physical parameters of a model.
Main observational signatures of the former values are gamma radiation and presence of antinuclei in cosmic rays. Indeed annihilations that take place at the borders between regions of matter and antimatter would contribute to a diffuse gamma-ray background. The distortion of CMB spectrum by released energy could be insignificant depends on the parameters of the antibaryon islands.

A close contact of coexisting matter and antimatter in the early epochs is almost unavoidable \cite{Kinney}. Equal number of matter and antimatter islands significantly contribute to the diffuse gamma-ray background \cite{Stecker,Cohen}. The gamma-ray flux in 100 MeV range caused by this kind of annihilation would be below the observable one only in the case when the characteristic size of domains exceeds $10^3$ Mpc. This fact requires baryon domination over the whole volume of the Universe.

Since antinuclei are very unlikely to be formed in the proton-antiproton (or proton-proton) collisions their presence in cosmic rays would be a strong indication of cosmic antimatter. Antinuclei could be naturally created inside the islands with high antibaryon density as a result of nucleosynthesis. Nowadays there are no evidences of antimatter domains existence \cite{Zitni, ARGO}. This does not exclude the case when the Universe is composed of matter with relatively small insertions of antimatter islands  \cite{Cohen}. The latter may be revealed as antistars distributed within our Galaxy \cite{Khlopov} or even distant galaxies. The absence of antihelium in the cosmic rays and annihilation signals leads to the conclusion that their fraction in the galaxy is smaller than $10^{-4}$. Analysis performed in \cite{Canetti} indicates that antimatter islands must be separated from a space filled with matter at least by the distance of about 1 Mpc. Nevertheless, some room even for antistars in our Galaxy still remains \cite{Vysotsky}.

Formation of antimatter islands of different sizes in the Universe is the subject of many baryogenesis models \cite{MatterAntimatter}. They all are based on the assumption of explicit C- and CP- violation. A detailed review of first models can be found in \cite{14}, \cite{Bernreuther}. The models of spontaneous baryogenesis are described in \cite{16}, \cite{Dolgov}, \cite{Takahashi}, \cite{Alberghi}. Another possible mechanism for antimatter domains origin is based on the class of Affleck-Dyne (AD) baryogenesis models. The foundation of AD baryogenesis is a peculiar evolution of the scalar field which carries the baryonic charge, on the background of cosmological expansion. This field is supposed to be coupled to supersymmetric particles, that store the baryon number when scalar field decays. The important feature of these models is the existence of ''flat directions'' in field space, in which the scalar potential vanishes \cite{Takahashi}, \cite{11}, \cite{13}, \cite{Takahashi2}, \cite{random3}. If some component of the complex scalar field moves along a flat direction, it can be considered as a free massless field, the so-called AD field. Usually this field behaves as Nambu-Goldstone (NG) boson.

The main aim of our paper is to show that the  antimatter domains can be manifested in the form of single galaxies with non-trivial chemical composition.

The proposed approach was based on the mechanism of spontaneous baryogenesis, and necessarily implied a complex scalar field carrying the baryonic charge. Necessary attribute of the previous paper \cite{Our} is the presence of large amplitude of initial baryon number fluctuations on the biggest cosmological scales, which leads to large amplitude of isocurvature fluctuations on large scales what contradicts the COBE data \cite{Burns}. One of the purpose  is to settle this issue and further develop the model.

The paper is organized as follows. In Sec. II we establish the model and discuss the quantum behavior of the scalar field in the inflationary period. Section III contains the size distribution of antimatter domains. In Section IV we discuss the observational manifestations. In Section V we conclude.

\section{Setup}

In this section we present a main tools of the model \cite{Our}
which is the basis of our study. Necessary corrections are also inserted to avoid the problem of the large scale fluctuations. The essence  of the problem is as follows. According to the model \cite{Our} antimatter domains of large scale are effectively produced due to fluctuations of some hypothetical scalar field. Unfortunately the amplitude of baryon fluctuations is too large what contradicts the COBE data (and the Planck data as well). To get rid of this problem one should suppress large amplitude fluctuations on the largest cosmological scales. In this paper we show that the effective suppression can be achieved by modifying the original potential used in paper \cite{Our}.
\begin{align} \label{potential}
V_0(\Psi)=\lambda\left(|\Psi|^2-\frac{f^2}{2}\right)^2+ \Lambda^4(1-cos(\theta)), \end{align}
where $\Psi = \chi(t)\exp(i\theta)$, $\chi(t)$ is a radial component of $\Psi$ field with minimum at $f/\sqrt{2}$ and $\theta$ is a massless NG field.  The first term looks similar to the Higgs potential which is known to be modified by quantum corrections \cite{Kamenshchik}. As the result it alternates its form depending on the top quark mass. Moreover landscape ideas \cite{Landscape}, \cite{random3}, \cite{random}, \cite{random1}, \cite{random2} imply potentials of arbitrary forms. Here we use this freedom to propose specific form of the potential
\begin{align} \label{potential2}
&V(\chi)=V_0(\chi) \cdot F(\chi), \\
F(\chi) = &\frac{C^2}{(|\chi|^a+C)^2}, \qquad \quad C, a = \text{const}>0  \nonumber
\end{align}
suitable for our purposes.
The modified potential \eqref{potential2} is close to the original one while $\chi$ is small.

During inflation, when the friction term is large the classical motion of the angular field $\theta$ is frosen due to the smallness of the potential tilt. The dynamics of the radial field $\chi$ is governed by the equation of motion
\begin{equation}\label{classeq}
\frac{\mathrm{d}^2 \chi(t)}{\mathrm{d} t^2}+3H\frac{\mathrm{d} \chi(t)}{\mathrm{d} t}+\frac{\mathrm{d} V\left(\chi\right)}{\mathrm{d} \chi} = 0
\end{equation}
where $H$ is the Hubble parameter.

While the field $\chi$ is moving classically to the potential minimum $\chi\simeq f/\sqrt{2}$, the phase $\theta=\varphi/f$ changes due to the quantum fluctuations against the de Sitter background \cite{21}, \cite{22}.
During inflation the amplitude of the phase $\theta$ changes as
\begin{equation}
\delta \theta \simeq \frac{H}{2\pi \cdot \overline{|\chi(t)|}}
\end{equation}
every e-fold, where  $\overline{|\chi(t)|}$ is the average modulus of the field $\chi$ during that e-fold.

The amplitude of baryon fluctuations $\delta{B_i}$ is proportional to $\delta \theta$ of the phase $\theta$ \cite{Liddle}, \cite{Our}. In the mentioned paper \cite{Our} $\chi(t)=f=const$ so that the amplitude fluctuations $\delta\theta$ are of the same order on different scales. This restriction is crucial for fitting the observational data. In our case the radial field $\chi(t)$ varies with time according to the classical equation \eqref{classeq} so that $\delta\theta$ could be small on large scales comparing to those on small scales.  This is the main difference between our model and the basic one \cite{Our}.

\begin{figure}[h]
\centering
\includegraphics[width=0.4\textwidth]{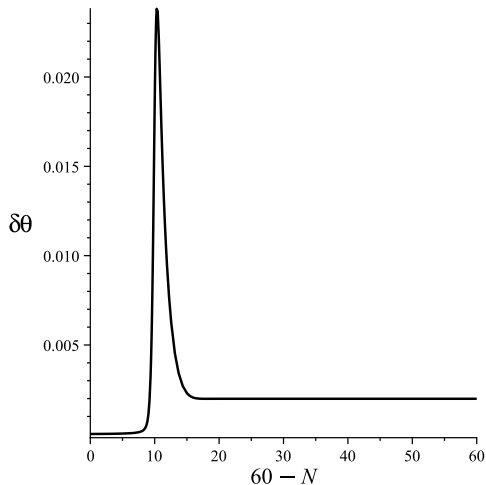}
\caption{Computed values of $\delta \theta$ as a function of (60-N), where $N$ is a number of e-folds, $N=H\cdot t$; The parameter values are: $f=80H$, $\lambda=8 \cdot 10^{-5}$, $C=4 \cdot 10^3 H^{a}$, $a=1.455$, $\chi_0 = 8.6 \cdot 10^3 H$}
\label{dTheta}
\end{figure}

Fig.\,\ref{dTheta} shows evolution of $\delta \theta$ in the modified potential. As one can see the phase fluctuations $\delta \theta$ remain small from the beginning and start growing after the $50$ e-fold when the largest scales have been formed.
Isocurvature fluctuations contribute to the CMB anisotropy as
\begin{equation} \label{fluct}
\frac{\delta T}{T} = \frac{1}{3}\frac{\Omega_B}{\Omega_0} \delta_{B_{i}}
\end{equation}
where $\delta_{B_{i}}$ is the amplitude of the initial baryon number fluctuations
and $\Omega_0 (\Omega_B)$ are the total (baryon) density in units
of critical density.
To avoid contradictions with COBE data one needs to have $\delta{B_i} \sim \delta \theta < 10^{-3}$ in first $6-8$ e-folds after the beginning of the inflation. With the mechanism proposed we have $\delta \theta \simeq 10^{-5}$. Hence the conflict with observations is settled.

The Planck data are satisfied as well. Recent results \cite{Planck2013} put a strong constraint on the isocurvature mode of fluctuations. The latter is characterized by the ratio \cite{Burns}
\begin{equation} \label{betaPlanck}
\beta_{iso} = \frac{1}{18\pi}\left( \frac{\Omega_B}{\Omega_0} \right)^2 \left( \frac{M_{Pl}}{|\chi| \theta_0}\right)^2 (-n_T)
\end{equation}

Since fluctuation arising in our mechanism is similar to the axion mode, we use constraints on axion isocurvature mode $\beta_{iso} < 0.039$ while $\beta_{iso} < 0.075$ in general case for the CDI component (CDM and baryon isocurvature mode) at low wavenumbers. The value of tensor spectral index $-n_T \ll 1$ and varies depending on an inflationary model. It is possible to satisfy constraints on $\beta_{iso}$ by choosing $-n_T$ in range $10^{-2} \div 10^{-4}$.

The baryon charge of the field $\Psi$ should be accumulated in matter fields. To make it possible, an interaction of the matter and the field $\Psi$ is postulated in the form

\begin{equation}\label{Lagrangian}
L_{int} = g \Psi \bar{Q} L + h.c.
\end{equation}
where fields $Q$ and $L$ represent a heavy quark and lepton fields coupled to the ordinary matter fields. We shall assume that fields $Q$ and $\Psi$ possess baryon number, while field $L$ does not. The U(1) symmetry that corresponds to the baryon number is expressed by following transformations:

\begin{align} \label{transformations}
\Psi \rightarrow \Psi \exp(i\beta) , \\
Q \rightarrow Q \exp(i\beta), \\ \nonumber
L \rightarrow L \nonumber
\end{align}

After the end of inflation the $\theta$ field starts to oscillate around zero with the frequency
\begin{equation}
\omega =m^{-1}_{\theta}=\frac{f}{\Lambda^2}
\end{equation}
Once oscillations begin, the energy density of the
PNG field converts to baryons and antibaryons.
The expression for baryon number
\begin{equation} \label{Bnumber}
N_{B(\bar{B})}\approx\frac{g^2f^2 m_{\theta}}{8\pi^2}\Omega_{\theta_i}\theta_{i}^{2}\int_{\mp \frac{\theta_{i}}{2}}^{\infty}\d \omega \frac{sin^2(\omega)}{\omega^2}
\end{equation}
can be found in \cite{Dolgov}, \cite{Our}, where $\Omega_{\theta_i}$ is a volume containing the phase value $\theta_i$ and coupling constant $g \ll 1$.

\section{Size distribution of antimatter domains}

In this Section we discuss the distribution of the antimatter domains following \cite{Our}. The total volume of all antimatter domains formed at the $N_t$\th e-fold before the end of inflation may be calculated using the recursive procedure: suppose the total volume of all the domains with the average phase $\bar{\theta}$ formed by that time is $V(\bar{\theta}, N_t)$. Then their total volume at the $(N_t-1)$\th e-fold before the end of the inflation is given by \cite{Our}
\begin{eqnarray}\label{iteration}
V(\bar\theta , N_{t}-1)=e^3\,V(\bar\theta ,N_t)+\Big[V_U(N_t)-e^3\,V(\bar\theta ,N_t)\Big]\cdot P(\bar\theta , N_{t}-1)\cdot h.
\end{eqnarray}
where $V_U(N_t)$ is the volume of the universe $V_U(N_t) \approx e^{3\cdot (N_U - N_t)}H^{-3}$ \quad $N_t$ e-folds before the end of inflation;\quad $N_U\approx 60$ is the total number of e-folds during inflation.  $P(\bar\theta , N_{t}-1)$ gives the Gaussian distribution of the phase \cite{21}:
\begin{align}
P(\bar{\theta}, N_t) = \frac{1}{\sqrt{2 \pi} \sigma_{N_t}} \cdot \exp\left( -\frac{(\theta_U-\bar{\theta})^2} {2 \sigma_{N_t}^2} \right), \\
\sigma_{N_t} = \frac{H}{2 \pi \cdot \overline{|\chi(N_t)|}} \cdot \sqrt{N_U-N_t}
\end{align}

Note that here $N_t$ is the number of e-folds \emph{before the end of inflation}, so it \emph{decreases} with time; so if the moment $N_0 \equiv N_U \approx 60, (t=0)$ corresponds to the beginning of inflation, while in the end of inflation $N_{\tau}=0$. Accordingly, the volume of antimatter domains in the beginning of inflation is $V(\bar{\theta},N_U) = 0$.

The first term in the equation \eqref{iteration} --  $e^3V(\bar\theta ,N_t)$ -- is the total volume of antimatter domains formed \emph{before} the $N_t$\th e-fold. The second term
\begin{equation}
v(\bar{\theta},N_t)=\Big[V_U(N_t)-e^3\,V(\bar\theta ,N_t)\Big]\cdot P(\bar\theta , N_{t}-1)\cdot h.
\end{equation}
is the total volume of antimatter domains formed \emph{during} the $N_t$\th e-fold. As the initial volume of each domain is $H^{-3}$, the number of domains formed during a given e-fold is
\begin{equation}
n = \frac{v(\bar{\theta}, N_t)}{H^{-3}}
\end{equation}
Domains grow in size during inflation, so the earliest domains to form become the biggest at the end of inflation. Their linear sizes at present time are defined by the equation
\begin{equation}
L(N_t) = 6\cdot10^3 e^{- (N_U - N_t)}
\end{equation}
Here $L(N_t)$ is the size in $Mpc$ of the antimatter domain formed at the $N_t$\th e-fold.

\begin{figure}[h]
\centering
\includegraphics[width=0.5\textwidth]{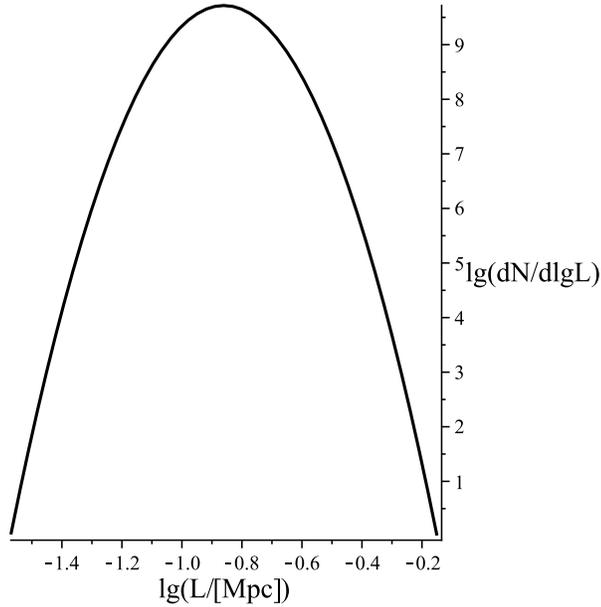}
\caption{Spectrum of antimatter domains, with parameters of the potential \eqref{potential2}: $f=80H$, $\lambda=8 \cdot 10^{-5}$, $C=4 \cdot 10^3 H^{a}$, $a=1.455$, $\chi_0 = 8.6 \cdot 10^3 H$, $\theta_0 = \frac{\pi}{15}$}
\label{Vol_distr}
\end{figure}

The result of numerical computations of the antimatter domains spectrum is represented in Fig.\,\ref{Vol_distr}. As one can see the maximum size of the largescale antimatter domains is about $~300$ kpc in the present epoch.

\section{Observational manifestations}

During the inflationary process antimatter domains of different size arise. If a domain survives till the temperature $~100$ keV  the nucleosynthesis starts inside it.
Antinuclei formation begins with the reaction
\begin{equation} \label{Breaction1}
\bar{p}+\bar{n} \rightarrow \bar{d} + \gamma
\end{equation}
This reaction strongly depends on an antibaryon density inside the domain.
For the reaction to be effective following condition has to be satisfied
\begin{equation} \label{Breaction2}
n_{\bar{B}}^{nucl} \langle \sigma \upsilon \rangle \geq H,
\end{equation}
where $\langle \sigma \upsilon \rangle = 4.55 \cdot 10^{-20}$ cm$^3$/sec \cite{Ours2} and one can use a relation $H = T^2/M_{pl}^{*}$ to go from the Hubble parameter $H$ to the temperature $T$.
Hence the anti-deuterium production takes place if the antibaryon density satisfy the inequality
\begin{equation} \label{Breaction4}
n_{\bar{b}}^{nucl} \geq 10^{18} \ cm^{-3}
\end{equation}

The value of $n_{\bar{b}}^{nucl}$ depends on the parameters such as $g$, $\Lambda$, $f$ and the value of $\theta$ in the certain domain.
Since this mechanism is responsible for baryon charge generation it has to produce appropriate amount of baryon asymmetry.
The observable number density of baryons is characterized by the ratio
\begin{equation}\label{excess}
\Delta_B = \frac{\Delta n_{B}}{s}= 0.86 \cdot 10^{-10}
\end{equation}
where $s$ - is the entropy density and $\Delta n_{B}$ - is the baryon excess.

According to \cite{Our}:
\begin{equation} \label{Basymmetry}
\Delta_B = \frac{45g^2}{16\pi^4g_{*}^{1/4}}\left(\frac{f}{M_{pl}}\right)^{3/2}\frac{f}{\Lambda}Y(\theta_0)
\end{equation}
where $Y(\theta)=\theta^2\int_{\frac{\theta}{2}}^{\frac{\theta}{2}}d\omega \frac{sin^2\omega}{\omega^2}$,$M_{pl}=10^6 H$, $g_{*}=106.75$.
It is easy to obtain the value $\Delta_B = 0.86 \cdot 10^{-10}$ varying parameters are $g$ and $\Lambda$ with selected values of $f$ and $\theta_0$.

\begin{figure}[h]
\centering
\includegraphics[width=0.6\textwidth]{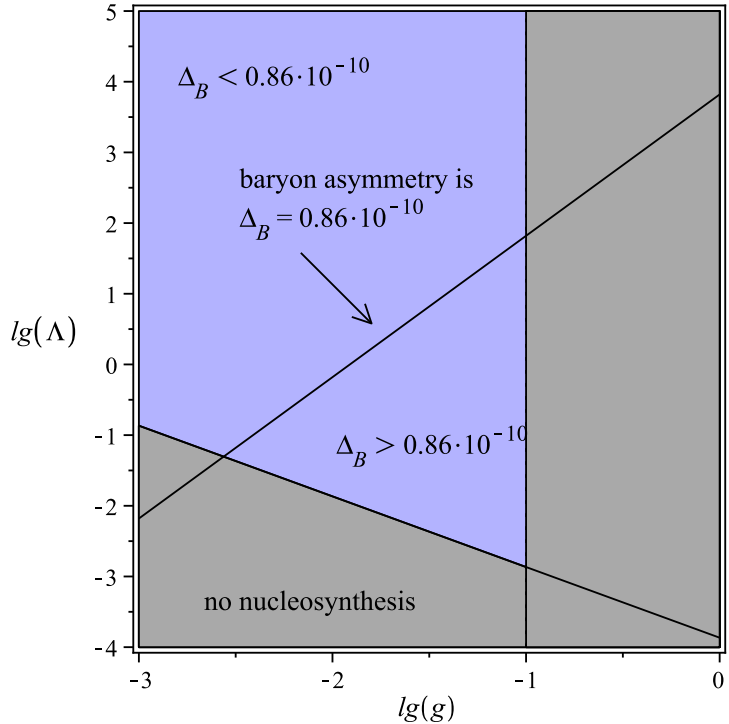}
\caption{Parameter space for the largest domains with biggest primordial antibaryon density.}
\label{Parameterspace}
\end{figure}

There is a wide room in the $g-\Lambda$ parameter space to create the conditions for nucleosynthesis to begin. Fig.\,\ref{Parameterspace} represents a $g-\Lambda$ parameter space for the largest domains that appeared at $50^{th}$ e-fold with the biggest primordial antibaryon density (since we are interested in the largescale antimatter domains that could possibly be conserved till present epoch). In the modern epoch the size of these domains is about $300$ kpc.

The right area and the area in the bottom left corner Fig.\,\ref{Parameterspace} are inscribed as "no nucleosynthesis" which means that with these values of $g$ and $\Lambda$ there will be no antideuterium production \eqref{Breaction1} in the largest domains. Preferred values (with nucleosynthesis going and $\Delta_B = 0.86 \cdot 10^{-10}$) of $g$ and $\Lambda$ are represented by the line. All the values that lie above the line will lead to the very small baryon asymmetry and the ones below the line lead to the excessive amount of baryon asymmetry.

The same parameter space plots could be drawn for every antimatter domain. Apparently there are too many parameters to take into account all of them, but we can do slices of parameter space as one presented in the Fig.\,\ref{Parameterspace}. Hereafter we will use $g=0.01$ and $\Lambda = 0.64$ to have baryosynthesis in most of largescale domains and satisfy \eqref{excess}. Note that there exist domains of various sizes (up to ~$300$ kpc) with $10^{-12} \leq \eta_{\bar{B}} \leq 6\cdot 10^{-10}$.

Due to the phase fluctuation, antibaryon density inside some domains can be very high in comparison with the background density. High density of antibaryons leads to the nontrivial chemical content after nucleosynthesis. Table.\,\ref{dens} shows the number of antimatter domains with densities exceeding the Universe average density. These regions contain more $^4 \bar{He}$ and $^7 \bar{Li}$, but the fraction of $\bar{D}$ and $^3 \bar{He}$ is reduced.

Note that this picture is similar to the Inhomogeneous Big Bang Nucleosynthesis (IBBN) \cite{Dolgov+}. In our case baryons are replaced by antibaryons and the difference between densities in different regions is not huge (IBBN model allow $\eta_{B} > 1$).

\begin{table}\label{dens}
\caption{Number of antimatter domains with very high densities.}
\begin{center}
\begin{tabular}{|c|c|c|c|c|c|}
$L$, kpc $\setminus \eta_{\bar{B}}$ & $7.8\cdot 10^{-10}$ & $9.4\cdot 10^{-10}$ & $1.1\cdot 10^{-9}$ & $1.4\cdot 10^{-9}$ & $1.6\cdot 10^{-9}$ \\
\hline
$100$ & $1.4 \cdot 10^5$ & $4.2\cdot 10^4$ & $9.2 \cdot 10^3 $ & $1.2\cdot 10^3$ & $90$ \\
\hline
$37$ & $1.6\cdot 10^3$ & $193$ & $13$ & $0$ & $0$ \\
\hline
$14$ & $214$ & $6$ & $0$ & $0$ & $0$ \\
\hline
$5$ & $66$ & $0$ & $0$ & $0$ & $0$ \\
\hline
$2$ & $29$ & $0$ & $0$ & $0$ & $0$ \\
\end{tabular}
\end{center}
\end{table}

Table.\,\ref{dens2} shows the abundances of antinuclei formed in the high-density antimatter domains. We used AlterBBN code (Version 1.4) \cite{AlterBBN} for numerical calculations.

\begin{table}\label{dens2}
\caption{Abundances of elements}
\begin{center}
\begin{tabular}{|c|c|c|c|c|c|c|}
$\eta_{\bar{B}}$ & $\bar{Y_p}$ & $\bar{^{2}H}/\bar{H}$ & $\bar{^{3}He}/\bar{H}$ & $\bar{^{7}Li}/\bar{H}$ & $\bar{^{6}Li}/\bar{H}$ & $\bar{^{7}Be}/\bar{H}$ \\
\hline
$7.8\cdot 10^{-10}$ & $2.497\cdot10^{-1}$ & $1.713\cdot10^{-5}$ & $8.894\cdot10^{-6}$ & $7.353\cdot10^{-10}$ & $7.625\cdot10^{-15}$ & $7.158\cdot10^{-10}$ \\
\hline
$9.4\cdot 10^{-10}$ & $2.514\cdot10^{-1}$ & $1.232\cdot10^{-5}$ & $8.050\cdot10^{-6}$ & $1.008\cdot10^{-9}$ & $5.564\cdot10^{-15}$ & $9.935\cdot10^{-10}$\\
\hline
$1.1\cdot 10^{-9}$ & $2.528\cdot10^{-1}$ & $9.157\cdot10^{-6}$ & $7.437\cdot10^{-6}$ & $1.273\cdot10^{-9}$ & $4.186\cdot10^{-15}$ & $1.262\cdot10^{-9}$\\
\hline
$1.4\cdot 10^{-9}$ & $2.550\cdot10^{-1}$ & $5.542\cdot10^{-6}$ & $6.639\cdot10^{-6}$ & $1.727\cdot10^{-9}$ & $2.582\cdot10^{-15}$ & $1.721\cdot10^{-9}$\\
\hline
$1.6\cdot 10^{-9}$ & $2.562\cdot10^{-1}$ & $4.065\cdot10^{-6}$ & $6.258\cdot10^{-6}$ & $1.999\cdot10^{-9}$ & $1.914\cdot10^{-15}$ & $1.995\cdot10^{-9}$\\
\end{tabular}
\end{center}
\end{table}
Upper limit to the ratio $\eta_{\bar{B}}$ is $1.1 \cdot10^{-9}$ due to the BBN constraints.
On the other hand no antinuclei are formed inside the regions with low antibaryon density and antiparticles remain ionized till recombination epoch.

This mechanism can lead to various antibaryon densities inside the domains of the same size.  The probability of low-density domain to appear is much higher than the one of high-density domain. Hence, low-density regions with $N_{\bar{B}} \approx 0$ are much more abundant so that high-density domains are mostly surrounded by a space with zero antibaryon charge. The size of these voids should not contradict CMB constraints.

According to our numerical calculations the largest scale of low-density antimatter regions is about $0.3$ Mpc at present
%
%
and there is no conflict with CMB. That differs from so-called Patchwork Universe. Matter-antimatter borders experience the lack of particles what reduces the annihilation rate in the epoch of inevitable annihilation \cite{Cohen}, \cite{Rujula}.

After the recombination large-scale structures begin to form. According to \cite{5}, high-density domains with masses exceeding the Jeans mass can evolve into compact stellar objects, globular clusters or even antigalaxies what also lower the annihilation signals. As a result we get compact antimatter object separated from the matter regions by low density voids.

The condition when the domain decouple from the cosmological expansion is
\begin{equation}
\label{jeans}
M_{\bar{B}} > M_J =\frac{4 \pi^{5/2}}{3}\upsilon^{3}_{s}\frac{M_{pl}^3}{\sqrt{\rho}}
\end{equation}
where $\rho$ is the background energy density and $\upsilon_s$ is the velocity of sound.
Antibaryon domains with masses $~10^{5}\div10^{6}$ M$_{\odot}$ are gravitationally unstable and could collapse to form bounded systems like GC or antigalaxies. Table\,\ref{galaxies} represents the number of antibaryon domains and their mass. That kind of a globular cluster (maybe with peculiar chemical composition) can exist in the nearby regions of our Galaxy.

\begin{table}{H} \label{galaxies}
\caption{Number of antibaryon domains and their mass}
\begin{center}
\begin{tabular}{|c|c|}
N & M/M$_{\odot}$\\
\hline
$511$ & $10^7$ \\
\hline
$2.3\cdot 10^6$ & $10^6$\\
\hline
$4.5\cdot 10^7$ & $10^5$ \\
\hline
$2\cdot 10^8$ & $10^4$\\
\end{tabular}
\end{center}
\end{table}

Along with large-scale domains that remain practically unaffected by the annihilation process there also exists a great number of "small" domains that could not survive till present time. Any antimatter regions with a size less than critical survival size $L_c \approx 1$ kpc in the contemporary epoch must be eaten up by the annihilation process \cite{Ours2}. Nevertheless it is important to show that these regions do not imply a significant distortion in the CMB spectrum.

Total amount of antibaryons within the domains must be small compared to the total baryon number of the Universe. As it follows from \eqref{Bnumber}, number of antibaryons depends on initial value of phase $\theta$ before the end of inflation. We suppose that the average step of phase is $\delta \theta = \frac{H}{2\pi f}$ at each e-fold, so that the phase value is distributed non-uniformly. Regions filled with phase $-\pi<\theta_i<0$ acquire nonzero antibaryonic charge. At the end of the inflation a size of regions is about $H^{-1}e^N$, where $N$ is a number of e-folds.
Under the parameters we use, the total number of antibaryons in all annihilated regions is $N_{\bar{B}} \approx 10^{29}$ what is negligibly small in comparison with a total number of baryons in the Universe. After its annihilation there will be no significant contribution to CMB.

Finally let us discuss the angular size of the antimatter domains and annihilation flux on Earth.
With new WMAP9 data it is possible to resolve objects of angular size of about $\Delta \alpha  \approx 0.1$. To remain hidden large antimatter domains of order $~30\div300$ kpc should be located at distances such that their angular size is less than $\Delta \alpha \approx 0.1$.
The latter is estimated as $\Delta \alpha = {L}/{l_d}$
where $L$ is a size of the domain, $l_d$ is an angular diameter distance.
Thus to mask their presence antimatter domains should be placed at $l_d > 300$ kpc depending on size ($l_d > 3$ Mpc for the largest domains of the size of $~300$ kpc ).

Gamma flux near Earth is expressed as \cite{Ours2}
\begin{equation} \label{Gammaflux}
\frac{d \Phi}{ d\Omega} \simeq 2.6 \cdot 10^{-4} \left( \frac{n_{\bar{b}}^{present}}{10^{-7} \; cm^{-3}} \right) \left( \frac{d}{l_d}\right)^2 \; cm^{-2} \; s^{-1} \; sr^{-1}.
\end{equation}
while the diffused photon background
is given by \cite{background}
\begin{equation} \label{Gammaflux2}
\frac{d \Phi}{ d\Omega} \simeq 10^{-8} \; cm^{-2} \; s^{-1} \; sr^{-1}
\end{equation}
below 1 GeV at high latitudes.
Therefore largest antimatter domain with highest antibaryon density should be positioned at the distance $l_d > 30$ kpc to stay unseen.
As the result the antimatter regions could be unseen even if they exist in form of globular clusters or diffused antimatter clouds near our Galaxy.

Recent discovery of antiproton signal at the energies about 100 GeV made by both PAMELA \cite{Galper} and AMS-02 \cite{AMS02} does not put a strong constraints on our model. The latter gives antiproton energies much smaller than 100 GeV and there is no realistic mechanism to increase it. Collisions of antiproton CR from nearby anti-galaxy with protons of galactic halos would result in multiple pion production due to reaction $p\overline{p}\rightarrow 5$ pions \cite{Pions}. Energy loss of pions and their decay give rise to diffuse gamma radiation with $E \approx 300$ MeV. Such an energy is close to lower threshold of Fermi LAT energy range.

Another observational signature would be interaction of anti-micrometeors of extra-galactic origin with the Sun and the Moon. Anti-meteors of different mass can be formed within anti-stellar systems. Collision with the Sun or the Moon would lead to the significant release of energy
\begin{equation}
E = 10^{18} \left(\frac{M}{1 mg} \right)\; erg .
\end{equation}
Annihilation rate limits can be found in \cite{KhlopovFargion}.

\section{Conclusion}

In the present paper we have developed a mechanism for antimatter region formation elaborated in \cite{Our}. The shortcoming of the previous model is the prediction of too large fluctuations of baryonic density, which contradicts observational data. The mechanism we propose removes that contradiction.

The model predicts the generation of large antimatter regions with sizes exceeding the critical surviving size. In the modern epoch domains with high  antimatter density evolve to single galaxies or globular clusters with a high content of antihelium and antideuterium. Such galaxies are separated from ordinary matter by voids of size at least ~$1$ Mpc, which is consistent with CMB data.

\section{Acknowledgment}
The work of A.G. and S.R. was supported by The Ministry of education and science of Russian Federation, project 3.472.2014/K.
We are very grateful for A.A. Kirillov and K.M. Belotsky for their interest in this research and fruitfull discussion.


\begin{thebibliography}{99}

\bibitem{ZeldDolg}
A.D. Dolgov, Ya.B. Zeldovich, Rev. Mod. Phys. 53 (1981)

\bibitem{ARGO}The ARGO-YBJ Collaboration,  	Physical Review D 85 (2012) 022002;	arXiv:1201.3848 [astro-ph.HE]

\bibitem{Zitni} A. Ariel Zhitnitsky, Phys.Rev. D74 (2006) 043515  arXiv:astro-ph/0603064 [pdf, ps, other]

\bibitem{MatterAntimatter}
A. Dolgov, Nucl.Phys.Proc.Suppl. 95 (2001) 42-46; A.D. Dolgov, Phys. Repts. 222, No. 6 (1992).

\bibitem{12}
R.W. Brown and F.W. Stecker, Phys. Rev. Lett. 43 (1979) 315.

\bibitem{14}
F.W. Stecker, Nucl. Phys. B252 (1985) 25.

\bibitem{Bernreuther}
W. Bernreuther, Lect.Notes Phys. 591 (2002) 237-293

\bibitem{16}
A. Cohen and D. Kaplan, Phys. Lett. 199B (1987) 251

\bibitem{Dolgov}
Alexandre Dolgov, Katherine Freese, Raghavan Rangarajan, Mark Srednicki, PHYSICAL REVIEW D 56, 10 (1997)

\bibitem{Takahashi}
F. Takahashi, M. Yamaguchi, Phys.Rev.D69:083506, (2004)

\bibitem{Alberghi}
G. L. Alberghi, R. Casadio, A. Tronconi, Mod.Phys.Lett.A22:339-346, (2007)

\bibitem{11}
I. Affleck and M. Dine, Nucl. Phys. B249, 361 (1985).

\bibitem{13}
M. Dine, L. Randall, and S. Thomas, Nucl. Phys. B458, 291 (1996).

\bibitem{Takahashi2}
Sh. Kasuya, F. Takahashi, Phys.Lett.B 736 (2014) 526-532

\bibitem{random3}
A.V. Grobov, S.G. Rubin, Phys.Lett. B726 (2013) 554-558

\bibitem{Stecker} F.W. Stecker and J.-L. Puget, Astrophys. J., 178, 57 (1972); G.A. Steigman, Annu. Rev.
Astron. \& Astrophys. 14, 339 (1976)

\bibitem{Cohen}
A.G. Cohen, A. De Rujula, and S.L. Glashow, Astrophys. J. 495, 539 (1998)

\bibitem{Kinney} W.H. Kinney, E.W. Kolb and M.S. Turner, Phys. Rev. Lett. 79, 2620 (1997)

\bibitem{Khlopov} A. Dudarewicz and A.W. Wolfendale, MNRAS 268 , 609 (1994);
M.Yu. Khlopov, Gravitation \& Cosmology, 4 , 69 (1998).

\bibitem{Our} M.Yu. Khlopov, S.G. Rubin, A.S. Sakharov, Possible origin of
antimatter regions in the baryon dominated universe, Phys. Rev. D 62, 083505 (2000); {http://arxiv.org/abs/hep-ph/0003285v1}{arXiv:hep-ph/0003285v1}

\bibitem{Burns}
S.D. Burns, astro-ph/9711303

\bibitem{Canetti}L. Canetti, M. Drewes, M. Shaposhnikov, arXiv:1204.4186v2

\bibitem{Kamenshchik}
A. O. Barvinsky, A. Yu. Kamenshchik, A. A. Starobinsky, JCAP 0811:021 (2008)

\bibitem{Vysotsky}A.D. Dolgov, V.A. Novikov and M.I. Vysotsky,
arXiv:1309.2746v2

\bibitem{Landscape}
Michael R. Douglas, JHEP 0305:046,2003

\bibitem{random}
S.G. Rubin,  Gravitation and Cosmology, 9 (2003) 243-248

\bibitem{random1}
S.G. Rubin,  arXiv:1403.2062

\bibitem{random2}
K.A. Bronnikov, S.G. Rubin, I.V. Svadkovsky, Phys.Rev.D 81, 084010, 2010

\bibitem{Planck2013}
Planck Collaboration, P. Ade et al., (2013), arXiv:1303.5082v2

\bibitem{21}
A. Starobinsky, Pis’ma Zh. Eksp. Teor. Fiz. 30, 719 (1979);
JETP Lett. 30, 682 (1979).

\bibitem{22}
A. Starobinsky, Phys. Lett. 91B, 99 (1980).

\bibitem{Ours2}
Evolution and observational signature of diffused antiworld
M.Yu. Khlopov, R.V. Konoplich, R. Mignani, S.G. Rubin,
A.S. Sakharov
Astroparticle Physics 12 2000. 367–372

\bibitem{Dolgov-obzor} A. D. Dolgov, Cosmological Matter-Antimatter asymmetry and Antimatter in the Universe, arXiv.org e-print archive, http://arxiv.org/abs/hep-ph/0211260, Accessed 19 July 2013.

\bibitem{5}
M.Yu. Khlopov, Gravitation and Cosmology, 4 (1998).

\bibitem{background}
Floyd W. Stecker, Tonia M. Venters, Astrophys. J. 736, 40 (2011)

\bibitem{Liddle}
D.H. Lyth and A. Riotto, Phys. Rep. 314, 1 (1999)

\bibitem{Rujula}
1997

\bibitem{ADolgov}
A. Dolgov, Matter-antimatter domains in the universe, (2001)

\bibitem{Dolgov+}
A.D. Dolgov, M. Kawasaki, N. Kevlishvili, Nucl.Phys.B807:229-250 (2009)

\bibitem{PBHclusters}
Vyacheslav Dokuchaev, Yury Eroshenko, Sergei Rubin, Grav.Cosmol. 11 (2005) 99-104

\bibitem{AlterBBN}
A. Arbey, Comput. Phys. Commun. 183 (2012) 1822, arXiv:1106.1363 [astro-ph.CO]

\bibitem{Galper}
O. Adriani et al., JETP Letters, (2012), Vol. 96, No. 10, pp. 621–627
 
\bibitem{AMS02}
https://indico.cern.ch/event/381134/contribution/2/material/slides/0.pdf

\bibitem{Pions}
Chechetkin et al., Rivista Nouvo Cim., 5, N-10,1 (1982);
Cohen et al., Astrophys. J. 495, 539 (1997)

\bibitem{KhlopovFargion}
M. Khlopov, D. Fargion, Astropart.Phys.19:441-446,2003
\end{thebibliography}
\end{document}